\documentclass[a4paper]{article}
\usepackage{Odyssey2024}
\usepackage{epsfig,amssymb,amsmath,booktabs}
\usepackage{cite}
\usepackage[hyphens]{url}
\usepackage{hyperref}
\usepackage[hyphenbreaks]{breakurl}
\usepackage{xcolor}
\usepackage{lipsum}
\usepackage{pifont,graphicx,adjustbox, comment}
\usepackage{multicol,multirow}
\usepackage{tabularx,adjustbox}
\usepackage{hyperref}
\usepackage{xspace}
\usepackage[subtle]{savetrees}
\hypersetup{
    colorlinks=true,
    linkcolor=purple,
    filecolor=magenta,      
    urlcolor=purple,
}
\usepackage{lipsum}

\ninept

\newcommand{\cmark}{\ding{51}}
\newcommand{\xmark}{\ding{55}}

\title{a-DCF: an architecture agnostic metric \\ with application to spoofing-robust speaker verification}

\name{\begin{tabular}{c}Hye-jin Shim,$^{1,*}$\thanks{*Equal contribution. Similar ideas were devised independently from two different groups and later combined.} Jee-weon Jung,$^{1,*}$
Tomi Kinnunen,$^{2}$ Nicholas Evans,$^{3}$
\\Jean-Francois Bonastre$^{4,5}$ and Itshak Lapidot$^{6,5,*\dag}$\thanks{$^\dag$ Corresponding author}\thanks{Code is available at \url{https://github.com/shimhz/a_dcf}.}\end{tabular}}
\address{$^1$Carnegie Mellon University, USA --
$^2$University of Eastern Finland, Finland\\
$^3$EURECOM, France -- 
$^4$Inria, France\\ 
$^5$Avignon University, France --
$^6$Afeka Tel-Aviv Academic College of Engineering, Israel
\\
\begin{tabular}{c}
{\small \tt shimhz6.6@gmail.com, jeeweonj@ieee.org, tomi.kinnunen@uef.fi,}\\ 
{\small \tt evans@eurecom.fr, jean-francois.bonastre@inria.fr, itshakl@afeka.ac.il}
\end{tabular}
} 
\begin{document}
\maketitle

\begin{abstract}

Spoofing detection is today a mainstream research topic.  
Standard metrics can be applied to evaluate the performance of isolated spoofing detection solutions and others have been proposed to support their evaluation when they are combined with speaker detection. 
These either have well-known deficiencies or restrict the architectural approach to combine speaker and spoof detectors. 
In this paper, we propose an architecture-agnostic detection cost function (a-DCF). 
A generalisation of the original DCF used widely for the assessment of automatic speaker verification (ASV), 
the a-DCF is designed for the evaluation of spoofing-robust ASV. Like the DCF, the a-DCF reflects the cost of decisions in a Bayes risk sense, with explicitly defined class priors and detection cost model.
We demonstrate the merit of the a-DCF through the benchmarking evaluation of architecturally-heterogeneous spoofing-robust ASV solutions. 

\end{abstract}

\section{Introduction}
%
%
%
All biometric verification systems, including automatic speaker verification (ASV), have the single task of determining reliably whether or not a biometric sample corresponds to a claimed identity~\cite{jain2004introduction}. 
Early on, reliability was interpreted as classifiers which could discriminate between target trials and (zero-effort)\footnote{{\it Zero-effort} implies purely casual impostors which make no concerted effort to deceive the system.} non-target trials~\cite{hansen2015speaker}.
In more recent years, and as a result of advances in text-to-speech synthesis and voice conversion, the threat of spoofing attacks, also known as presentation attacks~\cite{isostandard30107}, has come to the fore.

Despite the still-evolving consideration and study of spoofing, the single task of discriminating between target and non-target trials remains fundamentally unchanged; target trials should be accepted, while anything else should be rejected. 
The approach to address this challenging problem has nonetheless undergone rather more fundamental shifts.
While there are examples of alternative approaches even in the early literature~\cite{sizov2015joint, todisco2018integrated, gomez2020joint, li2020joint}, almost all studies of spoofing-robust ASV adopt the use of so-called {\it tandem} architectures~\cite{sahidullah2016integrated,shim2022baseline}.
These employ a pair of sub-systems, each also a binary classifier, one tasked with discriminating between target and non-target trials (the speaker detector), the other between bonafide and spoofed trials (the spoof detector).
The tandem approach is characteristic of the majority of related work, including studies involving other biometric traits~\cite{porwik2019ensemble,ergin2014ecg}. 

Standard metrics developed for the evaluation of speaker detectors can also be applied to the evaluation of spoof detectors, also known as countermeasures (CMs); they are both binary classifiers.
Alternative metrics proposed in recent years also support the evaluation of speaker and spoof detectors when combined~\cite{Kinnunen2020-tandem-fundamentals,kinnunen2023t}.
While the combination of speaker and spoof detectors still constitutes a single, binary classifier with the very same original task of accepting bonafide target trials and rejecting anything else, the consideration of spoofing complicates evaluation.
Despite still being a binary classifier, there are now three input classes ({\em target}, {\em non-target}, and {\em spoof}).
The tandem detection cost function (t-DCF)~\cite{Kinnunen2020-tandem-fundamentals} was hence developed to accommodate the evaluation of spoofing-robust ASV.

The t-DCF, however, is somewhat restrictive in terms of supported architectures. 
The speaker detector is used as a gate to the spoof detector,\footnote{There are plenty of other possible combination architectures.} while an \texttt{AND} decision logic is used to combine their respective classification decisions;
given a target trial, the spoof detector should indicate that an input utterance is bonafide \texttt{AND} the speaker detector should indicate that the input utterance corresponds to the claimed identity.
The t-DCF requires the computation of separate speaker and spoof detection scores and cannot be applied to the evaluation of alternative approaches which, for example, might produce only a single score.
Such approaches have been reported in the past and continue to emerge~\cite{jung2022sasv2,liu2024generalizing,gomez2020joint,li2019multi,shim2020integrated,zhangexplore,heo2022two}.
Their incompatibility with the t-DCF metric, has hence stimulated the development of alternative metrics, all forms of equal error rate (EER) estimates~\cite{jung2022sasv2}.
Different metrics are hence employed for the evaluation of different architectural approaches to the evaluation of spoofing-robust ASV solutions, despite them all sharing exactly the same goal.  
Use of different metrics then complicates the benchmarking of competing solutions. 
Furthermore, while an evaluation metric introduced in~\cite{greenberg2012nist} can be adapted to incorporate the aforementioned condition, no existing research explicitly defines this problem to date.

We aim to provide a solution to this problem.
We propose a derivative of the original DCF metric which is agnostic to the chosen architecture, be it a tandem system comprising separate, cascaded speaker and spoof detection sub-systems, alternatives whereby the roles of speaker and spoof detection are more closely integrated and perhaps jointly optimised, or indeed any other potential architecture.
The only demand of the architecture-agnostic detection cost function (a-DCF) is that, whatever the architecture might be, it must produce a single score which provides an indication of whether or not an input utterance corresponds to the claimed identity and is also bonafide.

\section{Related work}
\label{sec:related}
%
%
%
%

Spoof detectors were initially evaluated 
independently from speaker detectors, i.e.\ without consideration of impacts upon ASV behaviour.  There were a number of advantages to this approach. First, as a fledgling initiative, it seemed sensible to simplify the task so that researchers without expertise in ASV, but perhaps with other relevant expertise, could still participate in the development of spoof detection solutions.  Second, at the time the best architecture with which to combine speaker and spoof detectors was unclear. Third, without one having been identified, it seemed sensible to avoid the imposition of a particular combination architecture only for the sake of a common approach to evaluation.
Spoof detection performance was then estimated straightforwardly using an EER metric, a measure of discrimination between the two classes (bonafide and spoof).

Even if spoof detection solutions were initially evaluated independently, the potential for their interference with speaker detection is evident. 
Just like speaker detectors, spoof detectors make errors, either by accepting spoofs, or by rejecting bonafide trials; depending on the combination architecture, a speaker detector might not be able to recover from errors made by a spoof detector (or vice versa).  
The community was hence in need of a metric which could be used to evaluate the performance of \emph{combined} speaker and spoof detection systems. 


The tandem-detection cost function (t-DCF)~\cite{Kinnunen2020-tandem-fundamentals} was developed for this purpose.
Again, there were several advantages, not least because it provided an alternative to the EER metric. 
As a result of the assumed conditional independence of speaker and spoof detection decisions, the t-DCF also promotes the independent development of spoof detectors.\footnote{In principle, the conditional independence assumption of the two sub-systems even enables them to be developed using different datasets, yet they can still be evaluated in combination; in practice, this is not recommended, as it introduces an unnecessary data variation factor.} Decisions produced by a spoof detector can be combined with those made by a speaker detector to provide a measure of spoofing-robust ASV performance.  At the same time, use of a DCF-derived metric also helped bridge the gap between the relatively young spoof detection community and the mature speaker detection community which had embraced the DCF metric~\cite{DODDINGTON2000-NIST} decades earlier.

The well-known NIST-defined DCF~\cite{DODDINGTON2000-NIST}, given by
\begin{equation}
 \text{DCF} \left(t\right) := C_{\text{miss}} \pi_{\text{tar}} P_{\text{miss}}^\text{asv} \left(t\right) + C_{\text{fa}} \pi_{\text{non}} P_{\text{fa}}^\text{asv} \left(t\right), 
 \label{eq:dcf}
\end{equation}
reflects the cost of decisions in a Bayes risk sense~\cite{Jaynes03,DudaHartStork01} for an ASV system in the absence of spoofing attacks. In~\eqref{eq:dcf}, $\pi_\text{tar}$ and $\pi_\text{non}=1-\pi_\text{tar}$ are the class priors.  $C_\text{miss}$ and $C_\text{fa}$ are non-negative costs assigned to missed targets (false rejections) and non-target false alarms (false acceptance), respectively. $P_\text{miss}^\text{asv}(t)$ and $P_\text{fa}^\text{asv}(t)$ are the two ASV detection error rates as a function of the detection threshold, $t$.

The t-DCF metric can be used for the evaluation of tandem systems when the non-target class is augmented to include the potential for spoofing attacks. The general form of the t-DCF is \cite{Kinnunen2020-tandem-fundamentals}
\begin{align}\label{eq:tDCF-unconstrained}
    \text{t-DCF}(t_\text{cm}, t_\text{asv}) := &\; C_\text{miss}\cdot \pi_\text{tar} \cdot P_\text{miss,tar}^\text{tdm}(t_\text{cm},t_\text{asv})\\
         & + C_\text{fa,non} \cdot \pi_\text{non}\cdot P_\text{fa,non}^\text{tdm}(t_\text{cm}, t_\text{asv})\nonumber\\
         & +{\ } C_\text{fa,spf} \cdot \pi_\text{spf} \cdot P_\text{fa,spoof}^\text{tdm}(t_\text{cm}, t_\text{asv}),\nonumber
\end{align}
which now contains a third term related to false alarms stemming from spoofing attacks, and where the three \emph{tandem} (tdm) detection error rates are now functions of \emph{two} detection thresholds --- one for the spoof detector or CM sub-system ($t_\text{cm}$), one for the speaker detector or ASV sub-system ($t_\text{asv}$). Each of the tandem detection error rates are computed under the assumption that the speaker and spoof detectors make (class-conditionally) independent decisions. This leads to an \texttt{AND}-gate rule for combining the two sub-system decisions; refer to \cite[Section III.C]{Kinnunen2020-tandem-fundamentals} and \cite[Section 4.3]{kinnunen2023t} for details.

Computation of the t-DCF in~\eqref{eq:tDCF-unconstrained} requires two sets of detection scores, each corresponding to one of the two sub-systems. Nonetheless, one can choose to freeze either sub-system (including its operating point) to give a \emph{constrained} t-DCF which then becomes a function of a single detection threshold only. Specifically, the \emph{ASV}-constrained t-DCF~\cite[Eq.\ (10)]{Kinnunen2020-tandem-fundamentals} is given by 
\begin{equation}\label{eq:tdcf}
  \text{t-DCF}(t_\text{cm}) := C_0 + C_1\, P_\text{miss}^\text{cm}(t_\text{cm}) + C_2\, P_\text{fa}^\text{cm}(t_\text{cm}),
\end{equation}
where $P_\text{miss}^\text{cm}(t_\text{cm})$ and $P_\text{fa}^\text{cm}(t_\text{cm})$ are now the miss and false alarm rates for the spoof detector. The parameters $C_0$, $C_1$ and $C_2$ \cite[Eq.\ (11)]{Kinnunen2020-tandem-fundamentals} depend on pre-defined cost and prior parameters, in addition to the performance of a (frozen) ASV sub-system (speaker detector). 
The ASV-constrained t-DCF remains the cost of a complete (tandem) system --- but where the machine learning engineer's optimization efforts are constrained to modifying the spoof detector only, since both the ASV sub-system and the combination rule are already `written in stone'. For the ASVspoof challenges~\cite{2021asvspoof}, this kind of \emph{partial} optimisation strategy was adopted for the reasons noted above; the ASV system and $t_\text{asv}$ in~\eqref{eq:tDCF-unconstrained} were set by the organisers, while challenge participants could focus on improving the spoof detector, while having partial knowledge of speaker detector implementation.


Use of the t-DCF, whether \eqref{eq:tDCF-unconstrained} or \eqref{eq:tdcf}, is nowadays unnecessarily restrictive. It assumes a specific, non-customisable combination architecture whereby speaker and spoof detectors are used in cascade, and whereby decisions produced by each are combined with an \texttt{AND} decision logic. The constrained t-DCF \eqref{eq:tdcf} further restricts the class of spoofing-robust ASV models. Many other architectural approaches are also possible, for instance score and embedding level combination architectures --- or even systems comprising multiple spoof detection sub-systems. None of these systems can be evaluated using the t-DCF.  

\begin{table}[t]
  \caption{Three EERs used in the SASV 2022 challenge. System is targeted to accept ``+" trials and reject ``-" trials.}
  \centering
  \label{tab:three_eers}
  \begin{tabular}{lccc}
    \toprule
    & Target & Non-target & Spoof\\
    \toprule
    SV-EER & + & - &  \\
    SPF-EER & + &  & - \\
    SASV-EER & + & - & - \\
    \bottomrule
  \end{tabular}
\end{table}

An ensemble of three different metrics, all EER estimates, was developed subsequently to support the evaluation of more flexible spoofing-robust ASV architectures~\cite{jung2022sasv2}.  
Each one corresponds to an evaluation protocol comprising a different mix of trial types illustrated in~\autoref{tab:three_eers}. The traditional speaker verification EER (SV-EER) is used to measure the 
discrimination between targets an non-targets. The
spoofing EER (SPF-EER) measures the discrimination between bonafide targets and spoofs. 
Finally, the spoofing-aware speaker verification EER (SASV-EER)  measures discrimination between bonafide targets (which should be accepted) and everything else (bonafide non-targets and spoofing attacks, which should be rejected). 

The ensemble of metrics provides insights into classifier performance under different trial combinations, as well as the impacts of spoofing and spoof detection upon ASV (just like the t-DCF).  Even so, there are {\it three} metrics (instead of one), and they are all EER estimates (instead of DCF estimates). The shortcomings of \emph{any} EER-based metric are well-known~\cite{isostandard30107, isostandard19795, kinnunen2023t}. 
For ASV applications, one generally prioritises \emph{either} a low miss rate \emph{or} a low false alarm rate, which conflicts with the notion of \emph{equal} with the EER. Unlike 
DCF-based metrics,
the EER is not  
customizable or an optimization target for different applications. 
The SASV-EER metric which pools non-targets and spoofing attacks among the negative class
has even \emph{further} issues. The classic EER, when used to measure discrimination between any pair of \emph{two} classes is not dependent on the empirical class priors. 
As soon as the third class (spoofing attacks) is pooled with non-targets, 
however, this property no longer holds~\cite[Section 4.4]{kinnunen2023t}; the SASV-EER metric becomes a function of the empirical class priors. 
Hence, the metric itself depends on the evaluation data. 

Note the subtle, but critically important, difference in the role of the class priors for the DCF, t-DCF and the proposed a-DCF. For all these metrics, the class priors are \emph{not} a property of any dataset; they are \emph{not} estimated or computed from empirical data, but instead reflect the evaluator's assumed (uncertain) class priors for the given application. Different to the \emph{implicit} (hidden) weighting of the non-target and spoof false alarm rates for the SASV-EER, 
DCF-based metrics 
make the class priors (and consequences of classification errors) \emph{explicit}. For these reasons, \textbf{we strongly discourage any further use of the SASV-EER metric.} If one absolutely must report EERs, our recommendation is to limit such reporting to the SV-EER and SPF-EER. For the interested reader, we point to the recently-introduced \emph{concurrent tandem} EER (t-EER)~\cite{kinnunen2023t} metric which is not dependent on the empirical class priors. Nonetheless, similar to t-DCF, it can only be used to evaluate cascaded systems; and similar to the other EERs, it cannot be customised to different applications. 



It is time to reconsider the architectures for spoofing-robust ASV.  New metrics are needed for their comparative evaluation.  They should avoid the use of EER-based metrics, and especially use of the SASV-EER, and inherit the favourable properties of the DCF and t-DCF metrics with an explicit detection cost model.  Last, they should be agnostic to the classifier architecture.

\begin{table*}[t]
\caption{
Comparison on the characteristics of existing EER and DCF metrics used in the ASV and CM field. 
SASV refers to spoofing-robust ASV where ASV and CM sub-systems are combined to produce a single score output.
In the first row, `Explicit detection cost', `Independent of class prior' refers to whether the detection cost is explicitly used or not (difference between DCF and EER families) and whether it is affected by the number of different trial classes. `\# Supported classes' and `Scores required' indicate the number of class types considered and input score(s) needed for the calculation, respectively.
}
\label{tab:comparison}
\resizebox{\textwidth}{!}{
\begin{tabular}{l|c|c|c|l}
\toprule
& Explicit detection cost & Independent of class prior & \# Supported classes & \# scores required \\ \hline
DCF~\cite{DODDINGTON2000-NIST} & \cmark & \cmark & 2 & 1 \\
t-DCF~\cite{Kinnunen2020-tandem-fundamentals} & \cmark & \cmark & 3 & 2 (ASV and CM) \\
{\bf a-DCF} ({\it proposed}) & \cmark & \cmark & 3 & 1 (SASV) \\ 

\hline
EER & \xmark & \cmark & 2 & 1 (ASV or CM) \\
SASV-EER~\cite{jung2022sasv2} & \xmark & \xmark & 3 & 1 (SASV) \\
t-EER~\cite{kinnunen2023t} & \xmark & \cmark & 3 & 2 (ASV and CM)\\
\bottomrule
\end{tabular}
}
\end{table*}
\section{Architecture-agnostic DCF}

In this section we outline the theoretical basis for the proposed architecture-agnostic detection cost function (a-DCF).  We start with a general formulation (Section~\ref{ssec:general_form}), and show how it relates to the NIST-defined DCF in~\eqref{eq:dcf}, before presenting the specific formulation adopted for experiments reported later in the paper (Section~\ref{ssec:normalisation}).  We then present the a-DCF as a generalisation of the traditional two-class DCF (Section~\ref{ssec:generalization of NIST DCF}) before arguing how the a-DCF can be applied to the evaluation of other spoofing-robust ASV architectures.

\subsection{General form}
\label{ssec:general_form}

Let us assume a multi-class classification problem.
Let ${\cal A} = \left\{ {{a_1},{a_2}, \ldots ,{a_K}} \right\}$ be a set of $K$ ground-truth class labels, 
let $T \in \cal A$ a true class label for a given trial and $E \in \cal A$ is an estimated/predicted class label (classifier output). 
Let us define also the following:
\begin{itemize}
  \item $K \times 1$ column vector of class priors ${\pi_k} = \text{Pr}\left(  {{T = a_k}} \right) $:
      \begin{equation*}
        \Pi  = \begin{array}{*{20}{c}} {\left[ {\begin{array}{*{20}{c}} {{\pi _1}}&{{\pi _2}}& \cdots &{{\pi _K}} \end{array}} \right]}^{ \top } \end{array},
    \end{equation*}
    such that, ${\forall k, {\pi _k} \ge 0,\sum\nolimits_{k = 1}^K {{\pi _k}}  = 1}$.
  \item $K \times K$ matrix of classifier conditional probabilities:
        \begin{equation*}
        \begin{gathered}
        {\cal P} = \left[ {{P_{qk}} = \text{Pr}\left( {{E = a_q}\left| {{T = a_k}} \right.} \right)} \right] \\     
        {\begin{array}{*{20}{c}} {\forall q,k \in \left\{ {1, \ldots K} \right\},{P_{qk}} \ge 0,}&{\forall k,\sum\nolimits_{q = 1}^K {{P_{qk}}}  = 1}, \end{array}}
        \end{gathered}
    \end{equation*}

    where $P_{qk}$ is the conditional probability that the classifier outputs decision $a_q$ 
    given the ground-truth class $a_k$. 
  \item $K \times K$ matrix of conditional costs:
    \begin{equation*}
        \begin{gathered}
        {\cal C} =  \left[ {{c_{qk}}} \right] \\ {\begin{array}{*{20}{c}} {\forall q,k \in \left\{ {1, \ldots K} \right\}, {c_{qk}} \ge 0,}&{\forall k,{c_{kk}} = 0,} \end{array}}
        \end{gathered}
    \end{equation*}
where $c_{qk}$ is the cost of the classifier outputting decision $a_q$ given the ground-truth class $a_k$.  Without loss of generality, costs in $\cal C$ can be set to zero in the case of correct decisions ($c_{kk} = 0$), whereas incorrect decisions can be assigned non-negative, real values.
\end{itemize}

The {\it total} cost of making decisions can then be expressed in compact form by:

\begin{equation}
\label{eq:totalCost}
   \rm{C}_{T} = \textbf{1}_{1 \times K} \cdot \left( {\cal C} \circ {\cal P} \right) \cdot \Pi,
\end{equation}

\noindent where $\textbf{1}_{1 \times K}$ is a vector of ones which acts to sum the columns of the matrix resulting from the terms to the right, and where $ \circ $ is the Hadamard or entry-wise product. Though expressed using different formalism, ~\eqref{eq:totalCost} coincidences with ~\cite[Eq.\ (4)]{ferrer2022analysis}.

In practice, classifier conditional probabilities are defined for some operating point $t$ 
%
%
for which entries in the matrix of classifier conditional probabilities, now ${{\cal P} \left( t \right)}$, can be approximated by 
${p_{qk}}\left( t \right) \approx {\raise0.5ex\hbox{$\scriptstyle {{N_{qk}}\left( t \right)}$}
\kern-0.1em/\kern-0.15em
\lower0.25ex\hbox{$\scriptstyle {{N_k}}$}}$. 
$N_k$ is the number of trials belonging to class $a_k$ whereas
$N_{qk}\left( t \right)$ is the number of trials among them that are classified as belonging to class $a_q$.

\subsection{Spoofing-robust speaker verification}
\label{ssec:srsv}

For a standard ASV task, classifier decisions result in the labeling of an input trial as either a target or non-target, corresponding respectively to either accept or reject decisions.
In this case, there are two possible input classes and two possible classifier predictions ($K=2$).
The formulation in~\eqref{eq:totalCost} can then written as: 
\begin{equation}
 {\rm{C}_{T}}(t) = C_{\text{non,tar}} \pi_{\text{tar}} P_{\text{non,tar}} (t) + C_{\text{tar,non}} \pi_{\text{non}} P_{\text{tar,non}}(t), 
 \label{eq:dcfstart}
\end{equation}
where $c_{1,2}=C_{\text{non,tar}}$ and $c_{2,1}=C_{\text{tar,non}}$ are the costs of classifying targets as non-targets and non-targets as targets respectively, $\pi_{\text{tar}}$ and $\pi_{\text{non}}=1-\pi_{\text{tar}}$ are the class priors, and where $p_{\text{non,tar}}(t)$ and $p_{\text{tar,non}}(t)$ are the classifier conditional probabilities of each decision error at the set classifier threshold $t$.   
~\eqref{eq:dcfstart} is identical to the familiar detection cost function (DCF) in~\eqref{eq:dcf}
in which `miss' signifies targets mistaken for non-targets and `fa' (false alarm) signifies non-targets mistaken for targets.


\label{subsec: miltiNonTarget}

When subjected to spoofing attacks, the 
negative input class becomes a union of bonafide non-target (zero-effort impostor) trials and spoofed target trials.
For brevity, we refer to them simply as non-target and spoofed trials respectively. 
Different priors and costs can be assigned to each. 
Even so, the decision is still binary, with inputs still being labeled as either targets (positive class) or non-targets (negative class).   
Identical to the traditional ASV scenario (without spoofing attacks), there hence remains a single cost for missed target detections.
However, since there are two different negative class priors, one for non-targets and one for spoofs,\footnote{Note that positive and negative class assignments are arbitrary~\cite{Kinnunen2020-tandem-fundamentals}. In keeping with the literature, here we assume the assignment of bonafide target trials to the positive class and everything else to the negative class.}
there are likewise two (possibly different)\footnote{From an applications perspective, the mistaking of non-targets for targets when caused by an unintentional impostor may not have the same ramifications (cost) of when a spoof mistaken for a target is caused by a fraudster.  In the spirit of `an error is an error, no matter the cause', for all experiments reported later in the paper, we use the \emph{same} costs for each type of false alarm.} false alarm costs.
Classifier conditional probabilities can also be estimated in the usual way for a specified operating point or threshold $t$.

The priors, the matrix of classifier conditional probabilities and the matrix of conditional costs then have the form:
%
\begin{equation}
    \begin{array}{*{20}{c}}

{\Pi  = \left[ {\begin{array}{*{20}{c}}
{{\pi _{{\text{tar}}}}}\\
{{\pi _{{\text{non}}}}}\\
{{\pi _{{\text{spf}}}}}
\end{array}} \right]}\\ \\


{{\cal P}\left( t \right) = \left[ {\begin{array}{*{20}{c}}
{P_{{\text{tar,tar}}}\left( t \right)}&{P_{{\text{tar,non}}}\left( t \right)}&{P_{{\text{tar,spf}}}\left( t \right)}\\
{P_{{\text{non,tar}}}\left( t \right)}&{P_{{\text{non,non}}}\left( t \right)}&{P_{{\text{non,spf}}}\left( t \right)}\\
{P_{{\text{spf,tar}}}\left( t \right)}&{P_{{\text{spf,non}}}\left( t \right)}&{P_{{\text{spf,spf}}}\left( t \right)}
\end{array}} \right]}\\ \\


{\cal C = \left[ {\begin{array}{*{20}{c}}
0&{{C_{{\text{tar,non}}}}}&{{C_{{\text{tar,spf}}}}}\\
{{C_{{\text{non,tar}}}}}&0&0\\
{{C_{{\text{spf,tar}}}}}&0&0
\end{array}} \right]{\rm{ }}}
\nonumber
\end{array}
\end{equation}
%
%
The off-diagonal zero elements in the cost matrix correspond to the cost of mistaking non-target trials for spoofed trials and vice versa; the binary classifier distinguishes target trials from anything else and hence cannot distinguish between non-targets and spoofs.   
The total cost is then given again from~\eqref{eq:totalCost} by:
\begin{align}\label{eq:a-DCF-definition_pre}
    {\rm{C}_{T}}(t) &:= C_\text{non,tar}\pi_\text{tar}P_\text{non,tar}(t) \notag \\
    &\quad + C_\text{spf,tar}\pi_\text{tar}P_\text{spf,tar}(t) \notag \\    
    &\quad + C_\text{tar,non}\pi_\text{non}P_\text{tar,non}(t) \notag \\
    &\quad + C_\text{tar,spf}\pi_\text{spf}P_\text{tar,spf}(t).
\end{align}
Remembering that the negative class is the union of non-target and spoofed trials, and like the correspondence between~\eqref{eq:dcfstart} and~\eqref{eq:dcf}, we obtain the architecture-agnostic detection cost function (a-DCF):
\begin{align}\label{eq:a-DCF-definition}
    a\text{-}\text{DCF}(t) &:= C_\text{miss}\pi_\text{tar}P_\text{miss}(t) \notag \\
    &\quad + C_\text{fa,non}\pi_\text{non}P_\text{fa,non}(t) \notag \\
    &\quad + C_\text{fa,spf}\pi_\text{spf}P_\text{fa,spf}(t),
\end{align}
in which the first two lines of \eqref{eq:a-DCF-definition_pre} are reduced to a single term which encapsulates the miss-classification of target trials as non-targets.
$C_\text{miss}$, $C_\text{fa,non}$ and $C_\text{fa,spoof}$ are, respectively, the costs of missing (falsely rejecting) a target speaker, falsely accepting a non-target speaker, and falsely accepting a spoof. $\pi_\text{tar}$, $\pi_\text{non}$ and $\pi_\text{spoof}$ are the asserted class priors.\footnote{As a probability mass function, $\pi_\text{tar} + \pi_\text{non}+\pi_\text{spoof}=1$ and therefore any two of these priors are sufficient to characterize the class priors.} Finally, $P_\text{miss}(t)$ (the miss rate), $P_\text{fa,non}(t)$ (the non-target false alarm rate) and $P_\text{fa,spf}(t)$ (the spoof false alarm rate) are the respective empirical detection error rates at detection threshold $t$.\footnote{While, as argued above, costs $C_\text{fa,non}$ and $C_\text{fa,spf}$ can be set to different or identical values, empirical detection error rates $P_\text{fa,non}(t)$ and $P_\text{fa,spf}(t)$ are almost certainly different, with the latter normally being greater than the former.}

\subsection{Normalisation}
\label{ssec:normalisation}

Identical to the t-DCF, the value of a-DCF is not bounded and can be difficult to interpret. 
Therefore, 
similar to normalisation of the NIST DCF~\cite{DODDINGTON2000-NIST} as well as the t-DCF~\cite{Kinnunen2020-tandem-fundamentals}, we further scale the a-DCF using the cost of a \emph{default system}~\cite{Brummer2013-bosaris,Nautsch2019-PhDthesis} 
which is
configured to either accept ($t \rightarrow -\infty$) or reject ($t\rightarrow +\infty$) every trial:
    \begin{equation}
        a\text{-DCF}_\text{def} := \min \bigg\{C_\text{miss,tar}\pi_\text{tar}, C_\text{fa,non}\pi_\text{non}+C_\text{fa,spf}\pi_\text{spf}\bigg\}\nonumber
    \end{equation}
where the minimum costs of the two edge cases are considered (these expressions follow directly from the limit behavior of the miss and false alarm rates in~\eqref{eq:a-DCF-definition}).
The \emph{normalized} a-DCF is then given by:
    \begin{equation}
        a\text{-}\text{DCF}(t)_{\text{norm}}=\frac{a\text{-}\text{DCF}(t)}{a\text{-DCF}_\text{def}}.
        \label{eq:normDCF}
    \end{equation}
While in real operational settings the detection threshold $t$ must be set \emph{before} observing test data, for analysis purposes it is informative to report the lowest possible cost when $t$ is allowed to vary. 
The \emph{minimum} a-DCF is defined simply by:
    \begin{equation}
    a\text{-DCF}_\text{min}:=\min_{t \in \mathbb{R}} \   a\text{-DCF}_{\text{norm}}(t). 
    \label{eq:dcf_min}
    \end{equation}
While the normalised a-DCF function in \eqref{eq:normDCF} can exceed 1, which
implies that it performs worse than a non-informative system which either accepts or rejects every trial, the minimum is always bounded between 0 and 1. 
All detection costs reported later in the paper are normalised minimums computed according to \eqref{eq:dcf_min}.
For simplicity, they are denoted simply as a-DCF. 
 
\subsection{Relation to NIST DCF and t-DCF}
\label{ssec:generalization of NIST DCF}

It would be helpful to discuss the relation between a-DCF and other DCFs reported in~\cite{DODDINGTON2000-NIST,greenberg2012nist,Kinnunen2020-tandem-fundamentals}. 
The comparison is shown in \autoref{tab:comparison}.
The original DCF~\cite{DODDINGTON2000-NIST} in~\eqref{eq:dcf}, as endorsed by NIST through the speaker recognition evaluation (SRE) series, is designed for the benchmarking of binary classifiers which produce a single detection score --- such as standalone speaker or spoof detection systems. The NIST DCF can be seen as a degenerate case of the a-DCF with $\pi_\text{spf}=0$ (no spoofing attacks). Interestingly, the NIST 2012 SRE campaign \cite{greenberg2012nist} used a 3-term DCF similar to \eqref{eq:a-DCF-definition}, though the two FA rate terms are those of \emph{known/unknown nontargets} (as opposed to nontargets and spoofs as in our work). Though the 2012 DCF and $a$-DCF are in the same function class of evaluation metrics (i.e. differ only by the choice of costs and priors), they are motivated (and used) from entirely different perspectives.

The t-DCF~\cite{Kinnunen2020-tandem-fundamentals} in~\eqref{eq:tDCF-unconstrained} and~\eqref{eq:tdcf} is closely related to the a-DCF. The difference in terms of structure is that the t-DCF requires \emph{two} detection scores, as opposed to a single score in the case of the NIST DCF and the a-DCF. Nonetheless, by assigning a peculiar `dummy' countermeasure, we can view the a-DCF as a special case of the t-DCF. To be specific, by letting the threshold in~\cite[Eq. (7)]{Kinnunen2020-tandem-fundamentals} approach~$-\infty$, the spoof detection miss and false alarm rates approach 0 and 1, respectively. This yields precisely the a-DCF expression in \eqref{eq:a-DCF-definition}. Nonetheless, this is \emph{not} our point, as we discuss next. 
\begin{table}[!t]
\caption{Two $a$-DCF priors and costs accounting for different scenarios.}
\vspace{2pt}
  \centering
  \label{tab:cost_def}
  \resizebox{\columnwidth}{!}{
    \begin{tabular}{l|l|l|l|l|l|l}
    \toprule
     & $\pi_{\text{spf}}$ & $\pi_{\text{non}}$ & $\pi_{\text{tar}}$ & $C_{\text{miss}}$ & $C_{\text{fa,non}}$ & $C_{\text{fa,spf}}$ \\ \hline
    a-DCF1 & 0.05 & 0.01 & 0.94 &1 & 10 & 10\\ \hline
    a-DCF2 & 0.01 & 0.01 & 0.98 &1 & 10 & 10\\
    \bottomrule
    \end{tabular}}
\end{table}
\begin{table*}[!th]
\caption{
Results of cascade (decision-level tandem with an \texttt{AND} gate) systems.
The ASV system is a pre-trained RawNet3 model~\cite{jung2022pushing} whereas the CM is the AASIST or AASIST-L (a lightweight variant) model described in~\cite{jung2022aasist}.
The SASV, SV and SPF EERs are those defined in the context of the SASV 2022 challenge~\cite{jung2022sasv2}.  t-EER is the tandem EER~\cite{kinnunen2023t}. 
Results also reported for a-DCF and t-DCF metrics.
}
  \centering
  \label{tab:tandem_adcf}
  \resizebox{\textwidth}{!}{
    \begin{tabular}{c|l|cccc|cc|cc}
    \toprule
     & \textbf{Cascade system} & 
     \multicolumn{4}{c|}{\textbf{EERs}} & 
     \multicolumn{2}{c|}{\textbf{min a-DCFs}} & \multicolumn{2}{c}{\textbf{min t-DCFs}} \\ 
     & (threshold $t$) & \textbf{SASV-EER} & \textbf{SV-EER} & \textbf{SPF-EER} & \textbf{t-EER} & \textbf{a-DCF1} & \textbf{a-DCF2} & \textbf{t-DCF1} & \textbf{t-DCF2} \\ \hline
    \multirow{4}{*}{RawNet3 + AASIST} & ASV$\rightarrow$CM ($t_\text{asv}=0.05$) & 21.51 & 44.45 & 0.65 
    & \multirow{4}{*}{1.36}
    & 0.1492 & 0.4316 
    & 0.0226 &  0.0644\\
     & ASV$\rightarrow$CM ($t_\text{asv}=0.5$) & 0.89 & 0.74 & 0.91 
     &
     & 0.1492 & 0.0506 
     & 0.0358 & 0.0918\\
     & CM$\rightarrow$ASV ($t_\text{cm}=0.05$) & 1.10 & 0.69 & 1.49 
     &
     & 0.0240 & 0.0405 
     & N/A & N/A\\
     & CM$\rightarrow$ASV ($t_\text{cm}=0.5$) & 0.76 & 0.71 & 0.80 
     &
     & 0.0180 & 0.0382 
    & N/A & N/A\\
    \hline
    \multirow{4}{*}{RawNet3 + AASIST-L} & ASV$\rightarrow$CM ($t_\text{asv}=0.05$) & 21.68 & 44.71 & 0.82 
    & \multirow{4}{*}{1.23}
    & 0.1551 & 0.4405 
    & 0.0264 & 0.0633\\
     & ASV$\rightarrow$CM ($t_\text{asv}=0.5$) & 1.02 & 0.74 & 1.04 
     &
     & 0.0222 & 0.0560 
     & 0.0379 & 0.0883\\

     & CM$\rightarrow$ASV ($t_\text{cm}=0.05$)& 1.58 & 0.69 & 2.24 
     &
     & 0.0322 & 0.0453 
     & N/A & N/A \\
     & CM$\rightarrow$ASV ($t_\text{cm}=0.5$) & 1.02 & 0.78 & 1.30 
     &
     & 0.0241 & 0.0448 
     & N/A & N/A \\
     \bottomrule
    \end{tabular}
    }
\end{table*}

\subsection{a-DCF: beyond tandem systems}

While the above special cases reveal formal connections between cost functions, degenerate spoof prior and dummy spoof detectors are not in the \emph{typical scope} of conventional (spoofing-robust) ASV and tandem recognizers, respectively. The former aims to improve speaker discrimination, without any consideration for 
spoofing attacks, hence neither the benchmarking datasets, nor the evaluation metrics or even the systems need be concerned with the potential for
spoofing attacks. Likewise, the key feature of tandem systems is that they consist of two non-dummy sub-systems, the decisions of which result from the application of  \emph{finite} thresholds. A tandem system with either one of the two systems being a dummy system is not in the typical scope.

Consider now a slightly \emph{revised} tandem detection system which still uses separate thresholds for the speaker and spoof detector, but which outputs a single soft decision as long as the spoof detector score is high enough. Namely,
    \begin{equation}
        s = 
        \begin{cases}
            s_\text{asv}, & \text{if } s_\text{cm} \geq t_\text{cm}\\
            -\infty, & \text{if } s_\text{cm} < t_\text{cm},
        \end{cases}
    \end{equation}
where $s_\text{asv}, s_\text{cm} \in \mathbb{R}$ are the detection scores produced by the speaker detector (the ASV sub-system) and the spoof detector (the CM sub-system), respectively, and where $t_\text{cm}$ is a pre-set CM threshold. 




%
\section{Experimental setup}
While the key contribution of this work is the new a-DCF metric, we report also an example application to the evaluation of spoofing-robust ASV solutions. 
All experiments were performed using an open source, publicly-available implementation\footnote{\url{https://github.com/shimhz/a_dcf}}  which can be used to reproduce our results.

\subsection{Dataset}
We employ the ASVspoof 2019 logical access (LA) corpus~\cite{wang2020asvspoof} for all experiments.
The corpus is organised into three subsets: training, development, and evaluation.
The training and development sets include genuine and spoofed speech from 20 speakers (8 male, 12 female).
We employ the evaluation protocol\footnote{\url{https://github.com/sasv-challenge/SASVC2022_Baseline/blob/main/protocols/ASVspoof2019.LA.asv.eval.gi.trl.txt}} used for the SASV 2022 challenge~\cite{jung2022sasv2}.
The protocol was proposed initially for calculating the coefficients of ASV-constrained t-DCF (Eq. 10 in~\cite{Kinnunen2020-tandem-fundamentals}) in the ASVspoof 2019 challenge.
It consists of three trial types, {\em target}, {\em non-target}, and {\em spoof}, given an enrolment and test utterances. Hence it can be used for evaluating different types of ASV systems, including spoofing-robust ASV systems.

\subsection{Systems}
We consider three types of systems: cascade, jointly optimised, and single-model, to demonstrate the architecture-agnostic of the a-DCF metric.\footnote{Note that although we categorise systems in \autoref{tab:single_result} into three categories, it does not mean that all systems should fall into one of the three categories. `Jointly optimised' can be defined differently under diverse circumstances, where in our case, we use the term to call systems that jointly optimise pre-trained ASV and CM systems to derive the final score for each input.}
For the cascade (decision-level tandem) systems, RawNet3~\cite{jung2022pushing} is used as an ASV model.
As a CM system, AASIST~\cite{jung2022aasist}, AASIST-L~\cite{jung2022aasist}
are used and several cascade systems are composed.
In addition, we consider
systems submitted to the SASV 2022 challenge~\cite{jung2022sasv2,aleninid,wang2022dku,choi2022hyu} which explored diverse methodologies including score-level and embedding-level combinations.
For the single-model systems, we employ various versions from \cite{mun2023towards},
comprising MFA-Conformer~\cite{zhang2022mfa} and SKA-TDNN~\cite{mun2023frequency} models, where the training data varies throughout the devised multi-stage training.\footnote{S1, S2, S3, and S4 represent MFA-Conformer index 7 and SKA-TDNN indices 3, 7, and 11, respectively, from the model indices shown in \url{https://github.com/sasv-challenge/SASV2_Baseline}.}
To avoid confusion regarding system details and to focus on comparing metric measurements, all systems are denoted simply with the alphanumeric identifiers in \autoref{tab:single_result}.

\vspace{-3pt}
\section{Results}
\begin{table*}[t]
  \caption{
  EER, a-DCF, and t-DCF results for diverse spoofing-robust ASV systems.
  Results shown for four different single-model systems from~\cite{mun2023towards} (S1-S4) and three SASV 2022 challenge submissions (C1-C3). 
  C4 is the same RawNet3 + AASIST ($t_{\text{asv}}=0.5$) system for which results are also presented in \autoref{tab:tandem_adcf}. 
  The t-EER and t-DCF can be calculated only for system C4.  Tandem metrics require two separate scores, while EER and a-DCF metrics can be calculated for all systems.}
  \centering
  \label{tab:single_result}
  \begin{tabular}{l|l|cccc|cc|cc}
  \toprule
 & \multirow{2}{*}{\textbf{System type}} & \multicolumn{4}{c|}{\textbf{EERs}} & \multicolumn{2}{c|}{\textbf{min a-DCFs}} & \multicolumn{2}{c}{\textbf{min t-DCFs}} \\
 &  & \textbf{SASV-EER} & \textbf{SV-EER} & \textbf{SPF-EER} &  \textbf{t-EER} & \textbf{a-DCF1} & \textbf{a-DCF2} & \textbf{t-DCF1} & \textbf{t-DCF2}\\
 \toprule
 S1 & Single-model & 1.19 & 1.82 & 0.58  & N/A & 0.0222 & 0.0578 &N/A&N/A\\
S2 & Single-model & 1.25 & 1.27 & 1.23 & N/A & 0.0268 & 0.0417 &N/A&N/A\\
S3 & Single-model & 1.82 & 2.51 & 1.16 & N/A & 0.0366 & 0.0853 &N/A&N/A\\
S4 & Single-model & 2.48 & 3.32 & 1.56 & N/A & 0.0485 & 0.1068 &N/A&N/A\\
\hline
C1 & Jointly optimized (\em \small score) & 0.13 & 0.11 & 0.17 & N/A & 0.0032 & 0.0060 &N/A&N/A\\
C2 & Jointly optimized ({\em \small embedding}) & 0.28 & 0.28 & 0.28 & N/A & 0.0067 & 0.0147 &N/A&N/A\\
C3 & Jointly optimized (\em \small score) & 0.37 & 0.45 & 0.26 & N/A & 0.0080 & 0.0219 &N/A&N/A\\
C4 & Cascade & 0.89 & 0.74 & 0.91 & 1.36 & 0.1492 & 0.0506 & 0.0357 & 0.0358\\
\bottomrule
\end{tabular}

\end{table*}

The discussion below is oriented toward the comparison of different evaluation metrics, highlighting the merits and agnosticity of each, rather than the performance of different systems (which is not the objective).
Results are expressed in terms of the four different EERs, three from \autoref{tab:three_eers},\footnote{For the sake of completeness, we include the SASV-EER despite our previously stated reservations concerning its use.} the t-EER~\cite{kinnunen2023t}, the min t-DCF~\cite{Kinnunen2020-tandem-fundamentals}, where possible, and the a-DCF.

As illustrated in the second column of  \autoref{tab:tandem_adcf} which shows results only for cascade systems, results are shown for two different configurations in each case where either the speaker detector (an ASV sub-system), or the spoof detector (a CM sub-system) comes first.  The detection threshold for the first system is set to either 0.05 or 0.5.   t-DCF results are computed in the usual way.  The a-DCF results are computed according to \eqref{eq:dcf_min} in which the roles of the speaker and spoof detector can be interchanged. 
In both cases, the ASV sub-system is the RawNet3 model described in~\cite{jung2022pushing}.
The CM sub-system is either the AASIST or AASIST-L model, a lightweight variant, both described in~\cite{jung2022aasist}.
t-DCF1 and t-DCF2 results are computed with the same priors and costs as a-DCF1 and a-DCF2 respectively (see~\autoref{tab:cost_def}). 
Though different formulations are readily derived, computation of the t-DCF according to~\eqref{eq:tdcf} is specific to the evaluation of cascade systems comprising a spoof detector and a frozen speaker detector.\footnote{The unconstrained t-DCF  formulation in~\eqref{eq:tDCF-unconstrained} is more flexible and can be applied to the evaluation of other architectures, such as systems where the CM (rather than the ASV) subsystem is frozen. Nonetheless, this leads to different cost functions that are not comparable with~\eqref{eq:tdcf}, hence the omission of some results in~\autoref{tab:tandem_adcf}.} 
Finally, the t-EER~\cite{kinnunen2023t} is not dependent on the order of the two sub-systems, hence there is only one value for each combination of ASV and CM sub-systems.
%
%
SASV-EER, SV-EER and SPF-EER results 
can all be computed for all setups, but they remain EERs and suffer from the detractions described earlier. 
Use of the a-DCF avoids these issues with results also being computable for each setup, with a common formulation and for any given set of priors and costs. 

Shown in~\autoref{tab:single_result} are results for single-model, jointly optimised, and cascade systems. 
Four single models from~\cite{mun2023towards} are simply denoted as `S'. 
Three jointly optimised systems, namely the top-3 performing submissions~\cite{aleninid,wang2022dku,choi2022hyu} to the SASV 2022 challenge~\cite{jung2022sasv2}, and one cascade system from \autoref{tab:tandem_adcf} are denoted as `C'.
EERs, though similarly computable for all systems, are still EERs with the same shortcomings.  According to \eqref{eq:tdcf}, t-DCF values can be computed only for the cascade system and not for either single or jointly optimised systems. Once again, the a-DCF can be applied to the evaluation of all $8$ systems when configured to produce a single score, and with explicitly defined priors and costs. 

%

\section{Conclusions}
We propose an architecture-agnostic detection cost function (a-DCF) designed for the evaluation of spoofing-robust automatic speaker verification (ASV) systems. 
%
The a-DCF extends the time-tested DCF adopted by the ASV community decades ago.
An alternative to the derided equal error rate metric, the a-DCF reflects the cost of decisions in a Bayes risk sense, with explicitly defined class priors and detection cost model.  
The a-DCF is also more flexible than its previously proposed t-DCF cousin in terms of supported classifier architectures, so long as they can be configured to produce a single output score.
The a-DCF is also task-agnostic and could also be applied to studies involving other biometric traits, or indeed entirely different problems.

Experiments serve as a demonstration of the a-DCF in benchmarking a broad variety of competing spoofing-robust ASV solutions.
Even if developing the means to compare the performance of such different approaches was an objective of this work, these experiments are not, and were not intended to be, sufficient on their own to help identify the most promising. 
This requires far more extensive experimental analysis which is left for future work.


\section{Acknowledgement}
We appreciate Dr. Shinji Watanabe for the valuable discussions. The work has been partially supported by the Academy of Finland (Decision No. 349605, project ``SPEECHFAKES'') and by the Agence Nationale de la Recherche (ANR) in France (BRUEL project, No. ANR-22-CE39-0009). Experiments of this work used the Bridges2 system at PSC and Delta system at NCSA through allocations CIS210014 and IRI120008P from the Advanced Cyberinfrastructure Coordination Ecosystem: Services \& Support (ACCESS) program, supported by National Science Foundation grants \#2138259, \#2138286, \#2138307, \#2137603, and \#2138296.

\bibliographystyle{IEEEbib}
\bibliography{shortstrings,refs}

\end{document}